\documentclass[english,12pt]{article}
\usepackage{array}
\usepackage{graphicx}
\usepackage{amssymb}
\usepackage{amsmath}
\usepackage{multirow}
\usepackage{prettyref}
\usepackage{babel}
\usepackage{units}
\usepackage[latin1]{inputenc}
\usepackage{amsfonts}
\usepackage{amssymb}
\usepackage{babel}
\usepackage{cite}
\def\@fmsl@sh#1#2#3{\m@th\ooalign{$\hfil#1\mkern#2/\hfil$\crcr$#1#3$}}
 \def\eq#1\en{\begin{equation}#1\end{equation}}
\def\s[#1,#2]{[#1\stackrel{\star}{,}#2]}
\def\sx[#1,#2]{[#1\stackrel{\star_{x}}{,}#2]}

\newcommand{\nc}{\newcommand}
\nc{\beq}{\begin{equation}}
\nc{\eeq}{\end{equation}}
\nc{\beqa}{\begin{eqnarray}}
\nc{\eeqa}{\end{eqnarray}}

\def\bc{\begin{center}}
\def\ec{\end{center}}

\def\to{\rightarrow}

\def\gsim{\mathrel{\mathpalette\atversim>}}

\def\bc{\begin{center}}
\def\ec{\end{center}}

\def\gsim{\mathrel{\rlap{\lower4pt\hbox{\hskip1pt$\sim$}}

    \raise1pt\hbox{$>$}}}       

\def\gsim{\mathrel{\rlap{\lower4pt\hbox{\hskip1pt$\sim$}}
    \raise1pt\hbox{$>$}}}       

\begin{document}
\makeatletter
\def\fmslash{\@ifnextchar[{\fmsl@sh}{\fmsl@sh[0mu]}}
\def\fmsl@sh[#1]#2{%
  \mathchoice
    {\@fmsl@sh\displaystyle{#1}{#2}}%
    {\@fmsl@sh\textstyle{#1}{#2}}%
    {\@fmsl@sh\scriptstyle{#1}{#2}}%
    {\@fmsl@sh\scriptscriptstyle{#1}{#2}}}
\def\@fmsl@sh#1#2#3{\m@th\ooalign{$\hfil#1\mkern#2/\hfil$\crcr$#1#3$}}
\makeatother

\thispagestyle{empty}
\begin{titlepage}
\boldmath
\begin{center}
  \Large {\bf Non-Minimal Coupling of the Higgs Boson to Curvature in an Inflationary Universe}
    \end{center}
\unboldmath
\vspace{0.2cm}
\begin{center}
{  {\large Xavier Calmet}$^a$\footnote{x.calmet@sussex.ac.uk}, {\large  Iber\^ e Kuntz}$^a$\footnote{ibere.kuntz@sussex.ac.uk} {\large and} {\large Ian G. Moss}$^b$\footnote{ian.moss@newcastle.ac.uk} }
 \end{center}
\begin{center}
{\sl $^a$Department of Physics and Astronomy, 
University of Sussex, \\  Falmer, Brighton, BN1 9QH, U.K.
} \\
{\sl $^b$School of Mathematics and Statistics,  Newcastle University, \\ Newcastle Upon Tyne, NE1 7RU, U.K.}
\end{center}
\vspace{5cm}
\begin{abstract}
\noindent
In the absence of new physics around $10^{10}$ GeV, the electroweak vacuum is at best metastable. This represents a major challenge for high scale inflationary models as, during the early rapid expansion of the universe, it seems difficult to understand how the Higgs vacuum would not decay to the true lower vacuum of the theory with catastrophic consequences if inflation took place at a scale above $10^{10}$ GeV. In this paper we show that the non-minimal coupling of the Higgs boson to curvature could solve this problem by generating a direct coupling of the Higgs boson to the inflationary potential thereby stabilizing the electroweak vacuum. For specific values of the Higgs field initial condition and of its non-minimal coupling, inflation can drive the Higgs field to the electroweak vacuum quickly during inflation. 
\end{abstract}  
\end{titlepage}



\newpage
The non-minimal coupling $\xi \phi^2 R$ of scalars ($\phi$) to curvature $R$ has attracted much attention in the recent years. Indeed, in four space-time dimensions, $\xi$ is a dimensionless coupling constant and as such is likely to be a fundamental constant of nature. With the discovery of the Higgs boson, the only known fundamental scalar field so far observed,  it became clear that this parameter is relevant and should be considered when coupling the standard model of particle physics to general relativity. 

The value of the non-minimal coupling of the Higgs boson to curvature is a free parameter of the standard model of particle physics. There has been no direct measurement so far of this fundamental constant of nature.  The discovery of the Higgs boson at the Large Hadron Collider at CERN and the fact that the Higgs boson behaves as expected in the standard model implies that the non-minimal coupling is smaller than $2.6 \times 10^{15}$ \cite{Atkins:2012yn}.  This bound comes from the fact that for a large non-minimal coupling the Higgs boson would decouple of the standard model particles. We have little theoretical prejudice on the magnitude of this constant. Conformal invariance would require $\xi = 1/6$, but this symmetry is certainly not an exact symmetry of nature.

Assuming that the standard model is valid up to the Planck scale or some $10^{18}$ GeV, the early universe cosmology of the Higgs boson represents an interesting challenge. Given the mass of the Higgs boson which has been measured at 125 GeV and the current measurement of the top quark mass, the electroweak vacuum is at best metastable \cite{Degrassi:2012ry}. The implication of this metastability of the electroweak vacuum for the standard model coupled to an inflation sector has recently been discussed \cite{Lebedev:2012sy}. Indeed, one finds that the Higgs quadratic coupling which governs the shape of the Higgs potential for large field value turns negative at an energy scale $\Lambda\sim10^{10}-10^{14}$ GeV. The electroweak vacuum with the minimum at 246 GeV is not the ground state of the standard model, but rather there is a lower minimum to the left and our vacuum is only metastable. This is a problem in an inflationary universe.

In an expanding universe with Hubble scale $H$, the evolution of the Higgs boson $h$ is given by
\begin{eqnarray}
\ddot h + 3 H \dot h + \frac{\partial V(h)}{\partial h}=0
\end{eqnarray}
where $V(h)$ is the potential of the scalar field. Even if one imposes as an initial condition at the start of our universe that the Higgs field starts at the origin, it will most likely be excited to higher field values during inflation. Indeed, because the mass of the Higgs boson is very small compared to the scale of inflation, it is essentially massless.  Quantum fluctuations of the Higgs field will drive it away from the minimum of the potential. Its quantum fluctuations are of order the Hubble scale $H$. Thus, for $H > \Lambda$, it is likely that the Higgs will overshoot the barrier between the false vacuum in which our universe lives and the lower state true vacuum of the theory.

In \cite{Lebedev:2012sy,Lebedev:2012zw}, it is shown that a direct  coupling of the Higgs boson to the inflaton field can significantly affect this picture if this coupling makes the Higgs potential convex. This interaction between the inflaton and the Higgs boson drives the Higgs field to small values during inflation. This is closely related to an earlier claim \cite{Espinosa:2007qp} that the curvature coupling of the Higgs boson resembles an additional mass term $-\xi R$ in the Higgs potential and could stabilise the Higgs boson. We shall argue below
this interpretation of the curvature term is not entirely correct, and in fact the two mechanisms are closely related when carried out correctly.
Assuming that there is no new physics between the weak scale and the scale of inflation, we shall derive a new prediction for the value of the non-minimal coupling of the Higgs boson to the Ricci scalar.

Before the discovery of the Higgs boson, cosmologists had already been investigating the non-minimal coupling of scalars to curvature. In inflationary cosmology one often deals with actions of the type
\begin{equation}\label{nonMin}
S_{scalar}=\int d^4x \,\sqrt{-g} \, \left (\frac{1}{2}\partial_\mu \phi \partial^\mu \phi - \frac{1}{2}m^2 \phi^2 + \frac{1}{2}\xi \phi^2 R \right),
\end{equation}
where $m$ is the mass of the scalar field $\phi$. This coupling has been extensively studied, see e.g. \cite{Chernikov:1968zm,Callan:1970ze,Frommert:1996gu,CervantesCota:1995tz,Bezrukov:2007ep}. With the discovery of the Higgs boson,  it became clear that this coupling was not only an exotic term that could be implemented in curved space-time but that this coupling is phenomenologically relevant. 

Before deriving our prediction for the value of the non-minimal coupling of the Higgs boson to curvature, we need to address a common misconception which can be very important when discussing Higgs physics within the context of cosmology and very early universe physics.  It is often argued that the non-minimal coupling which appears in Eq.(\ref{nonMin}) of a scalar field to curvature is identical to a contribution to the mass of the scalar field that is curvature dependent. We will prove that this is not strictly correct. We will then show that the non-minimal coupling of the Higgs boson to curvature does actually help to stabilize the Higgs potential, and furthermore it can even
drive the Higgs field towards the false vacuum from a Planck-scale initial value.

We shall first address the issue of the Higgs mass. If one naively varies the action for a scalar field $\phi$ containing the non-minimal coupling (\ref{nonMin}), one obtains the field equation
\begin{eqnarray}
(\Box + m^2 - \xi R)\phi=0,
\end{eqnarray}
and it is often argued that this term $\xi R$ is a curvature dependent mass term for the scalar field $\phi$. In a FRW background,
the curvature drops from $R=12H^2$ during inflation, with constant expansion rate $H$, to $R\approx 0$ in a radiation dominated era
after inflation, which could lead to an overproduction of the Higgs boson after inflation \cite{Herranen:2015ima}. This argument is however incomplete. The problem is that the non-minimal coupling  induces a mixing between the kinetic term of the scalar field and of the metric field. We will illustrate this point with the standard model of particle physics, since this is the only model so far that contains a fundamental scalar field which has actually been discovered experimentally, however the same line of reasoning applies to any scalar field non-minimally coupled to curvature.

Starting with the standard model Lagrangian $\mathcal{L}_{SM}$, we have
\begin{eqnarray}
\label{action2}
S = \int d^4x \, \sqrt{-g} \left[ \left( \frac12  M^2 + \xi {\cal H}^\dagger {\cal H} \right) R - 
(D^\mu {\cal H})^\dagger (D_\mu {\cal H}) - \mathcal{L}_{SM}  \right] 
\end{eqnarray}
where ${\cal H}$ is the SU(2) scalar doublet, we shall see that this is not actually the Higgs boson of the standard model. After electroweak symmetry breaking, the scalar boson gains a non-zero vacuum expectation value, $v=246$ GeV, $M$ and $\xi$ are then fixed by the relation
\begin{eqnarray}
\label{effPlanck}(M^2+\xi v^2)=M_P^2 \, .
\end{eqnarray}

The easiest way to see that ${\cal H}$ is not actually the Higgs boson is by doing a transformation to the Einstein frame \cite{vanderBij:1993hx,Zee:1978wi,Minkowski:1977aj} $\tilde g_{\mu\nu}=\Omega^2 g_{\mu\nu}$, where $\Omega^2 =(M^2+2\xi {\cal H}^\dagger {\cal H})/M_P^2$. The action in the Einstein frame then reads
\begin{eqnarray}
\label{actionEins}
S=\int d^4x \, \sqrt{- \tilde g} \left[\frac{1}{2}M_P^2 \mathcal{\tilde R} -
\frac{3 \xi^2}{M_P^2 \Omega^4}\partial^\mu ({\cal H}^\dagger {\cal H} ) \partial_\mu 
({\cal H}^\dagger {\cal H} ) -\frac{1}{\Omega^2} (D^\mu {\cal H})^\dagger (D_\mu {\cal H}) - 
\frac{\mathcal{L}_{SM}}{\Omega^4}  \right] \, .\end{eqnarray}
Expanding around the Higgs boson's vacuum expectation value and specializing to unitary gauge, 
${\cal H}=\frac{1}{\sqrt{2}}(0,\phi + v)^\top$, we see that in order to have a canonically normalized kinetic term for the physical Higgs boson we need to transform to a new field $\chi$ where
\begin{equation}
\label{chiphi}
\frac{d\chi}{d\phi}=\sqrt{\frac{1}{\Omega^2}+\frac{6\xi^2 v^2}{M_P^2 \Omega^4}}\, .
\end{equation}
Expanding $1/\Omega$, we see at leading order the field redefinition simply has the effect of a wave function renormalization of $\phi = \chi/\sqrt{1+\beta}$ where $\beta =6 \xi^2 v^2 / M_P^2$.  Thus the canonically normalized scalar field, i.e., the true Higgs boson, does not have any special coupling to gravity and it couples like any other field to gravity in accordance with the equivalence principle.

This effect can also be seen in the Jordan frame action (\ref{action2}) as arising from a mixing between the kinetic terms of the Higgs and gravity sectors. After fully expanding the Higgs boson around its vacuum expectation value and also the metric around a fixed background, $g_{\mu\nu}=\bar \gamma_{\mu\nu}+h_{\mu\nu}$, we find a term proportional to $\xi v \phi \square h^\mu_\mu$:
\begin{eqnarray}
{\cal L}^{(2)}&=&-\frac{M^2 +\xi v^2}{8}(h^{\mu\nu}\Box h_{\mu\nu}+ 2 \partial_\nu h^{\mu\nu} \partial_\rho h^{\mu\rho} 
-2 \partial_\nu h^{\mu\nu} \partial_\mu h^{\rho}_{\rho}
-h^{\mu}_{\mu}\Box h^{\nu}_{\nu} \\ \nonumber &&
+\frac{1}{2} (\partial_\mu \phi)^2 + \xi v ( \Box h^{\mu}_{\mu}-\partial_\mu \partial_\nu h^{\mu\nu} )\phi
\end{eqnarray}
After correctly diagonalizing the kinetic terms and canonically normalizing the Higgs field and graviton using
\begin{eqnarray}
\phi&=&\chi/\sqrt{1+\beta} \\
h_{\mu\nu}&=&\frac{1}{M_P} \tilde h_{\mu\nu} - \frac{2 \xi v}{M_P^2 \sqrt{1+\beta}}\bar \gamma_{\mu\nu}\chi.
\end{eqnarray}
We again find the physical Higgs boson gets renormalized by a factor $1/\sqrt{1+\beta}$.

These results demonstrate that the non-minimal coupling does not introduce stronger gravitational interactions for the Higgs boson once its field has been correctly canonically normalized. We stress that the underlying reason is that there is no violation of the equivalence principle. Our findings are in sharp contrast to the claims made in \cite{Herranen:2014cua}. The only valid bound to date on the non-minimal coupling of the Higgs boson to curvature is that obtained in \cite{Atkins:2012yn}, namely that its non-minimal coupling is smaller than $2.6\times10^{15}$.  While the fact that we may be living in a metastable vacuum is problematic for the Higgs boson in an inflationary context, the non-minimal coupling of the Higgs boson to curvature does not create a new problem. On the contrary, we shall now show that this non-minimal coupling could solve the stability issue.

Let us now study the coupling of the Higgs boson to an inflationary potential $V_I(\sigma)$ that is induced by the mapping from the Jordan frame to the Einstein frame. Indeed, even if no direct coupling between the Higgs boson is assumed in the Jordan frame, it will be induced in the Einstein frame:
\begin{equation}
 V_I(\sigma) \to  \frac{V_I(\sigma \Omega)}{\Omega^4}=
  \frac{V_I(\sigma \Omega)}{\left ( 1+ \frac{2\xi v\phi(\chi)+\xi\phi(\chi)^2}{M_P^2}\right)^2},
 \end{equation}
 but bear in mind that the inflaton field $\sigma$ does not have a canonically normalized kinetic term. 
 
 Let us first consider Higgs field values $v\ll\phi \ll M_P|\xi|^{-1/2}$). In that case, we see immediately that
 \begin{equation} \label{eqinfpot}
 \frac{V_I(\sigma\Omega)}{\Omega^4}\approx V_I(\sigma)\left(1-2\xi\phi^2/M_p^2\right).
 \end{equation}
A coupling between the inflaton and the Higgs field is induced by the transformation to the Einstein frame. Note that there is a priori no reason to exclude a coupling of the type $V_I {\cal H}^\dagger {\cal H}$ in the Jordan frame where the theory is defined. There could be cancelations between this coupling and that generated by the map to the Einstein frame. The magnitude of the coupling between the Higgs boson and the inflaton appearing in the mapped inflationary potential thus cannot be regarded as a prediction of the model. Let us ignore a potential direct inflaton-Higgs coupling for the time being and continue our investigation of the induced coupling. We will now show that a non-minimal coupling of the Higgs boson to curvature can solve some of the problems associated with Higgs cosmology within the standard model of particle physics.

In the early universe we need to consider large Higgs field values ($\phi\gg v$). As explained previously, even if one is willing to fine-tune the initial condition for the value of the Higgs field, it will experience quantum fluctuations of the order of the Hubble scale $H$. Unless the Hubble scale is much smaller  than the energy scale at which the electroweak vacuum becomes unstable, the Higgs field is likely to swing into the lower true vacuum of the theory.  A Higgs non-minimal coupling to the Ricci scalar could actually solve this problem since, as we will show, it will generate a direct coupling between the Higgs boson and the inflaton if the Jordan frame action contains an inflationary potential $V_I$. 

It has been shown that a direct coupling between the Higgs boson and the inflaton can drive the Higgs field \cite{Lebedev:2012sy} to the false electroweak vacuum  quickly during inflation even if the Higgs field initial value is chosen to be large. There are basically three scenarios for the onset of inflation: the thermal initial state \cite{guth}, ab initio creation \cite{vilenkin,hawking} and the chaotic initial state \cite{linde2,linde3}. The thermal initial state starting from a temperature just below the Planck scale  would introduce thermal corrections to the Higgs potential preventing vacuum decay until the temperature fell to the inflationary de Sitter temperature, at which point it becomes a question of vacuum fluctuation as to whether the Higgs survives in the false vacuum. However, the consistency of thermal equilibrium of the standard model fields when the Higgs takes a large value has not yet been verified. The ab initio creation is an attractive possibility, where the Higgs would nucleate at the top of the potential barrier. In this case also, stability depends on the size of vacuum fluctuations during inflation. The final possibility, the chaotic initial state, would have the Higgs field start out at arbitrarily large values. The most likely initial values would be larger than the instability scale 
$\Lambda$, preventing the Higgs field from entering the false vacuum. An anthropic argument could be applied to rule out these initial conditions, but we shall see that the non-minimal curvature coupling of the Higgs boson can force the Higgs into the false vacuum without anthropic considerations.

\begin{figure}[htb]
\begin{center}
\includegraphics[width=0.5\textwidth]{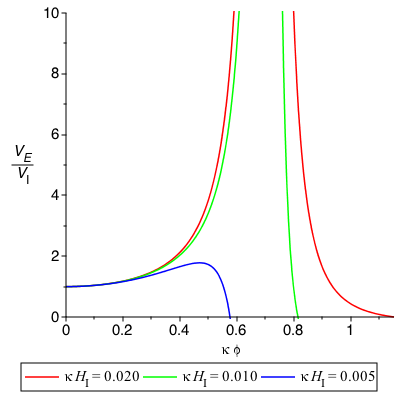}
\caption{
The Einstein frame Higgs potential $V_E(\phi)$ for different values of the false-vacuum
inflation rate $H_I$ for $\xi=-2$. The potential vanishes at $\phi=\phi_m$, and there
is an asymptote at $\phi=\phi_c$. Consistency of the model (no ghosts) requires $\phi<\phi_c$. 
An initial condition $V_E\sim M_p^4$ can be achieved with the initial $\phi$
close to $\phi_c$.
}
\label{fig:pot}
\end{center}
\end{figure}

As we have seen, the Einstein frame potential is given by
\begin{equation}
V_E={V_I(\sigma)+V_\phi(\phi)\over (1+\xi\kappa^2\phi^2)^2}
\end{equation}
where $\kappa^2=8\pi G$. The inflationary expansion rate $H_I$ is the expansion rate of the false vacuum,
\begin{equation}
H_I^2={V_I(\sigma)\over 3 M_p^2}.\label{hi}
\end{equation}
The most extreme chaotic initial condition, and the one relevant to eternal chaotic inflation,
is one where $V_E$ is close to the Planck scale. For an unstable Higgs potential $V_\phi$, 
this is only possible when $\xi<0$, as shown in Fig. \ref{fig:pot}.  

Let us denote by $\phi_m$ the value of the field at which the potential vanishes,
\begin{equation}
V_I(\sigma)+V_\phi(\phi_m)=0.\label{defphim}
\end{equation}
Note that $\phi_m$ depends on $H_I$. The asymptote in the potential is at $\phi_c$,
\begin{equation}
1+\xi\phi_c^2/M_p^2=0.
\end{equation}
Provided that $\phi_c<\phi_m$, then there is an initial value of $\phi$ close to $\phi_m$
at which $V_E\sim M_p^4$ (note that it has been shown in \cite{Calmet:2013hia} that even with a large non-minimal coupling of the Higgs boson to curvature, the cutoff of the effective field theory can be as large as the Planck scale), since $\phi=\phi_c$ is an asymptote. If $\phi_c>\phi_m$, then there is no such value.

Starting form the initial value, the Higgs field evolves to small field values on a timescale 
comparable to the Hubble expansion rate. Unfortunately, we cannot simply expand the conformal factor in the denominator of the Einstein frame potential for all values of $\xi$. However, it is straightforward to see this effect  from kinetic terms of the Higgs boson and of the inflaton. The kinetic terms for the Higgs and inflaton are multiplied by $g_\phi$ and $g_\sigma$ respectively, where
\begin{equation}
g_\phi={1+\xi\kappa^2\phi^2+6\xi^2\kappa^2\phi^2\over (1+\xi\kappa^2\phi^2)^2},\qquad 
g_\sigma={1\over (1+\xi\kappa^2\phi^2)^2}
\end{equation}
Note that is it possible to use a canonically normalised Higgs field $\chi$ as we had done previously, but not
both the Higgs and inflaton fields at the same time because the field space metric is curved.

The early evolution of the Higgs field is described by the equation
\begin{equation}
\ddot\chi+3H\dot\chi+{dV_E\over d\chi}=0.
\end{equation}
For the inflaton, one has
\begin{equation}
(g_\sigma\dot\sigma)\dot{}+3Hg_\sigma\dot\sigma+{dV_E\over d \sigma}=0,
\end{equation}
while the expansion rate is given by
\begin{equation}
3H^2=\kappa^2\left(\frac12g_\sigma\dot\sigma^2+\frac12\dot\chi^2+V_E\right).
\end{equation}
The inflaton equation can also be written as
\begin{equation}
\ddot\sigma+\left({1\over g_\sigma}{dg_\sigma \over d\chi}\right)\dot\chi\dot\sigma
+3H\dot\sigma+{1\over g_\sigma}{dV_E\over d \sigma}=0.\label{sigmaeq}
\end{equation}
Note that the second term in this equation is not considered in \cite{Lebedev:2012sy}. For $\chi>M_p$, we have
\begin{equation}
V_E\approx(V_I+V_\phi)e^{\sqrt{8/3}\kappa(\chi-\chi_0)},\qquad
g_\sigma\approx e^{\sqrt{8/3}\kappa(\chi-\chi_0)}.
\end{equation}
There is thus rapid evolution of $\chi$ and slow evolution of $\sigma$ (assuming slow-roll
conditions on $V_I$). Indeed, the inflaton evolves 
on a longer timescale than the Higgs field, leaving a gradual reduction in $H_I$, and also $\phi_m$. 
Eventually the potential
evolves to $\phi_c>\phi_m$, but at all stages the Higgs field lies on the false vacuum
side of the potential barrier. As long as the vacuum fluctuations do not cause quantum
tunnelling, the Higgs field will enter the false vacuum.

The condition that $\phi_c<\phi_m$ implies limits on the curvature coupling $\xi$.
In order to determine these limits we need to calculate $\phi_m$ from (\ref{defphim}),
and this requires an expression 
for the Higgs potential.
For a standard model Higgs field, the large field Higgs potential in flat space is given by
\begin{equation}
V_\phi=\frac14\lambda(\phi)\phi^4
\end{equation}
In curved space, the Higgs develops a mass of order $H$ multiplied by Higgs couplings,
but we can think of this as a radiative correction to $\xi$ and regard $\xi$ as the
effective curvature coupling at the inflationary scale. 
Other curvature corrections to the Higgs potential
may well be important, but for now these will be neglected.

The effective Higgs coupling $\lambda(\phi)$ vanishes at some large value of $\phi$
which we identify as the instability scale $\Lambda$. The value of $\Lambda$ is very
strongly dependent on the top quark mass, and currently all we can say is that it
lies in the range $10^9-10^{18}$ GeV. Furthermore, adding additional particles to the
standard model changes the instability scale (or removes the instability altogether).
It is therefore convenient to give results treating $\Lambda$ as a free parameter.
In the range of Higgs field values where the potential barrier lies, we
use an approximation to the running coupling given by 
\begin{equation}
\lambda(\phi)\approx b\left\{\left(\ln{\phi\over M_p}\right)^4-\left(\ln{\Lambda\over M_p}\right)^4\right\},
\end{equation}
with $b\approx 0.75\times 10^{-7}$. This fits quite well to the renormalisation group calculations
\cite{Degrassi:2012ry}.

\begin{figure}[htb]
\begin{center}
\includegraphics[width=0.5\textwidth]{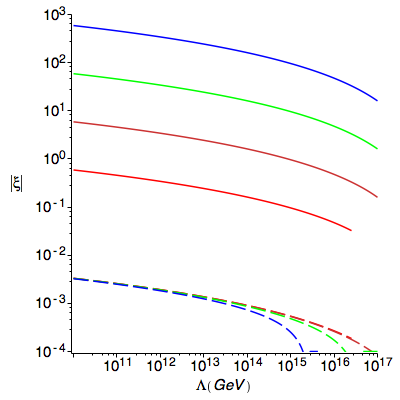}
\caption{
The lower bound on $-\xi$, where $\xi$ is the curvature coupling, for consistent chaotic initial conditions
on the Higgs field which will lead the Higgs into the false vacuum. The horizontal
axis is the Higgs stability scale. The different curves
from bottom to top are for the false vacuum Hubble parameter $0.1M_p$ to $10^{-4}M_p$.
The dashed lines show the lower bound for quantum stability of the false vacuum.
}
\label{fig:pot2}
\end{center}
\end{figure}

The plots in Fig. (2) show numerical results for the values of $-\xi$ which are lower bounds
of the range which is consistent with chaotic initial conditions. 
Also shown by the dashed lines are the quantum bounds from the vacuum tunnelling
rate $\exp(-8\pi^2\Delta V_E/3 H_I^2) \sim O(1)$, where $\Delta V_E$ is the height of the
potential barrier \cite{hawking}. 
(The quantum bound on $-\xi$ is lower than the one quoted in \cite{Herranen:2014cua},
which we believe is due to our inclusion of the $8\pi^2/3$ factor.)
The results show curves for different values of the false vacuum Hubble parameter,
essentially corresponding to different initial values of the inflaton field through (\ref{hi}). We ought
to expect that this initial Hubble parameter is close to the Planck scale. As advertised, a non-minimal coupling of the Higgs boson can drive the Higgs boson into the false vacuum of the standard model early on during inflation. Instead of being a source of problems, it can solve some of the issues associated with the cosmological evolution of the Higgs boson.

It is worth mentioning as well that our results also imply that the non-minimal coupling of the Higgs boson will not influence reheating as long as the Higgs field value is small during inflation. Reheating could be generated by a direct coupling of the Higgs boson to the inflaton via either couplings of the type $\sigma^2 {\cal H}^\dagger {\cal H}$ or $\sigma {\cal H}^\dagger {\cal H}$. As usual right-handed neutrinos $N$ could also play a role in reheating via a coupling $\bar N N \sigma$.However, none of these couplings will be significantly influenced by the conformal factor or the rescaling of the Higgs boson as long as one is considering small Higgs field values. 

We have seen that a non-minimal coupling of the Higgs boson to the Ricci scalar does not generate new issues for Higgs boson physics in the early universe and that, on the contrary, there is a range of values for $\xi$ for which the Higgs potential is stabilized thanks to the coupling of the Higgs boson to the inflaton generated by the non-minimal coupling of the Higgs boson to curvature. This becomes obvious when mapping the Jordan frame action to the Einstein frame. Finally, it has been shown in \cite{Calmet:2013hia}  that the non-minimal coupling $\xi$ does not introduce a new scale below the Planck mass which finishes to establish our point that the standard model, if we add a non-minimal coupling to the Ricci scalar, could be valid up the Planck scale in an inflationary universe.

\noindent{\it Acknowledgments:}
This work is supported in part by the Science and Technology Facilities Council (grant numbers  ST/L000504/1
and ST/J000426/1) and by the National Council for Scientific and Technological Development (CNPq - Brazil).


\bigskip{}

\end{document}